\documentclass[twocolumn]{aastex631}

\usepackage{graphicx}	% Including figure files
\usepackage{amsmath}	% Advanced maths commands
\usepackage{subfigure}
\usepackage{appendix}

\begin{document}
% \linenumbers 
\title{Dynamic disk–corona coupling during the state transition of Swift J1727.8–1613}

\author[0009-0007-7292-8392]{Han He}
\affiliation{Department of Astronomy, School of Physics and Technology, Wuhan University, Wuhan 430072, People's Republic of China}

\author[0009-0007-1966-181X]{Yi Long}
\affiliation{Department of Astronomy, Nanjing University, 163 Xianlin Avenue, Nanjing 210023, People's Republic of China}

\correspondingauthor{Bei You}

\author[0000-0002-8231-063X]{Bei You}
\affiliation{Department of Astronomy, School of Physics and Technology, Wuhan University, Wuhan 430072, People's Republic of China}
\email{youbei@whu.edu.cn}

\author[0000-0001-9969-2091]{Fu-Guo Xie}
\affiliation{Shanghai Astronomical Observatory, Chinese Academy of Sciences (CAS), Shanghai 200030, People's Republic of China}

\author[0000-0002-5385-9586]{Zhen Yan}
\affiliation{Shanghai Astronomical Observatory, Chinese Academy of Sciences (CAS), Shanghai 200030, People's Republic of China}

\author[0000-0002-0333-2452]{Andrzej A. Zdziarski}
\affiliation{Nicolaus Copernicus Astronomical Center, Polish Academy of Sciences, Bartycka 18, PL-00-716 Warszawa, Poland}

\author[0009-0007-6828-3931]{Sai-En Xu}
\affiliation{Department of Astronomy, School of Physics and Technology, Wuhan University, Wuhan 430072, People's Republic of China}

\newcommand\hh[1]{{\color{magenta}Han: #1}}
\newcommand\ly[1]{{\color{blue}Yi: #1}}
\newcommand\xse[1]{{\color{orange}Saien: #1}}
\newcommand\by[1]{{\color{red}BY: #1}}

\begin{abstract}

State transitions during outbursts of black hole X-ray binaries exhibit complex, rapidly evolving disk–corona coupling. Understanding this dynamic phase is essential for deciphering accretion physics and the mechanisms that drive outbursts, yet it remains poorly understood because of the scarcity of high-quality, high-cadence observations. Here, we present an analysis of observations from the Hard X-ray Modulation Telescope (HXMT) during the 2023 outburst of the newly discovered low-mass black hole X-ray binary Swift J1727.8–1613. Follow-up, high-cadence monitoring reveals pronounced variability in disk emission, attributable to fluctuations in the accretion rate. These disk fluctuations exhibit damped amplitudes and shortened flare periods. This evolving disk emission modulates the supply of soft seed photons to the corona, producing a dynamically changing positive correlation between the photon index $\Gamma$ and the Comptonization luminosity $L_{\rm Comp}$. As the transition proceeds, the correlation shifts toward higher $\Gamma$ and a narrower range of $L_{\rm Comp}$. We further suggest that the damped disk variability arises from fluctuations generated at large disk radii and propagating inward, possibly linked to the thermal–viscous disk instability.

\end{abstract}

\section{Introduction}
\label{sec:intro}

Black hole X-ray binaries (BHXBs) are among the most powerful laboratories for studying accretion physics, exhibiting dramatic spectral and timing transitions over months to years \citep{Remillard2006ARA&A..44...49R,Done2007A&ARv..15....1D,Belloni2016ASSL..440...61B}. In the hard state, the X-ray emission is dominated by Comptonization in a hot corona or an advection-dominated accretion flow (ADAF) \citep{Narayan1994ApJ...428L..13N,Yuan2014ARA&A..52..529Y}, accompanied by strong broadband variability and low-frequency quasi-periodic oscillations (QPOs) \citep{Belloni2016ASSL..440...61B,Ingram2019NewAR}. As the source brightens and transitions to the soft state, the multi-temperature blackbody component from an optically thick, geometrically thin accretion disk becomes dominant \citep{Shakura1973A&A....24..337S}, while high-frequency variability is typically suppressed \citep{Ingram2019NewAR}.

The transition between the hard and soft states, on the timescale of days to months, is closely tracked by the photon index \citep{Belloni2016ASSL..440...61B}, which encodes the dynamic coupling between the disk and corona or ADAF \citep{Beloborodov1999ASPC..161..295B,Done2007A&ARv..15....1D,Liu2022iSci...25j3544L}. Specifically, the hard-to-soft transition is characterized by progressive spectral softening as the photon index $\Gamma$ increases, a decline in the high-energy Comptonized flux, a brightening and hardening of the thermal disk component, and a suppression of broadband timing variability \citep{Done2007A&ARv..15....1D,Belloni2016ASSL..440...61B}.

Physically, the driving mechanism behind state transitions remains debated. The truncated disk model proposes that, in the hard state, the inner disk recedes to a truncation radius of tens of gravitational radii and is replaced by a hot, radiatively inefficient flow. As the source softens, the disk refills toward the innermost stable circular orbit (ISCO) \citep{Esin1997ApJ...489..865E, Done2007A&ARv..15....1D, you2023b, you2023Sci,liu2026}. The disk-corona evaporation model describes mass exchange between the thin disk and the overlying hot corona via thermal conduction, with the disk refilling and the corona condensing as the accretion rate rises \citep{Meyer2000A&A...361..175M, Qiao2013ApJ...764....2Q, Liu2022iSci...25j3544L}. The jet-disk coupling model links the evolution of the compact jet, which dominates the radio emission in the hard state, to changes in the disk-corona geometry, with the jet quenching coinciding with the transition to the soft state \citep{Fender2004MNRAS.355.1105F, Fender2009MNRAS.396.1370F,Corbel2013MNRAS.428.2500C}. In addition, Lense-Thirring precession is invoked to explain quasi-periodic modulations and the onset of state transitions in misaligned systems \citep{Nixon2014MNRAS.437.3994N}. Despite these theoretical advances, characterizing disk-corona activity during a state transition remains challenging because most sources are monitored with limited cadence. Here, we analyze the 2023 outburst of the black hole X-ray binary Swift J1727.8-1613, which was observed with extensive HXMT data during its exceptionally long ($\sim$43-day) hard-to-soft state transition, providing a rare opportunity to study the dynamic disk--corona coupling in detail throughout the transition.

Swift J1727.8-1613 is a newly identified low-mass black hole X-ray binary, first detected by Swift/BAT on August 24, 2023 (MJD 60180; \citealt{Negoro2023ATel}). Follow-up observations were conducted with a sequence of instruments, including NICER, NuSTAR, and MAXI \citep{peng2024ApJ...960L..17P}. Recent optical observations derived a distance of $d=3.4\pm0.3\,\text{kpc}$ (\citealt{Mata2025A&A...693A.129M}; but see also \citealt{Burridge2025ApJ...994..243B,Zdziarski2025ApJ...986L..35Z}). Dynamic analysis established a lower limit for the compact object's mass of $M_1>3.12\pm0.10\rm ~M_{\odot}$ and refined the orbital period to $P_{\rm orb}=0.45017\pm 0.00004~\rm d$ \citep{Mata2025A&A...693A.129M}. In addition, the hard X-ray modulation telescope (HXMT) conducted long-term, high-cadence follow-up observations from August 25, 2023 (MJD 60181) to October 6, 2023 (MJD 60223), during which Swift J1727.8-1613 displayed a peak flux of 7 Crab in the 15-50 keV range, making it one of the brightest X-ray binary systems \citep{Palmer2023ATel16215....1P, Zhao2024ApJ...961L..42Z}. Based on spectral and timing analyses, Swift J1727.8-1613 remained in the state transition for 43 days, from September 2, 2023 (MJD 60189) to October 15, 2023 (MJD 60232) \citep{Katoch2023ATel16235, Katoch2023ATel16243, Mereminskiy2023arXiv, Yu2024MNRAS.529.4624Y, Trushkin2023ATel16289....1T, mata2024A&A...682L...1M}. This was corroborated by the Rms-Intensity Diagram (RID; see Section \ref{subsec:hid&rid}). During this prolonged state transition, HXMT recorded multiple flares in the low-energy band and a relatively smooth decline in the high-energy band \citep{Yu2024MNRAS.529.4624Y}, indicating strong activity in the accretion disk \citep{cao2025}.

In this paper, we investigate the pronounced X-ray flux variability observed in Swift J1727.8-1613 during the outburst. The X-ray observations from HXMT and the associated data reduction are described in Section \ref{sec:data}. We present our principal findings in Section \ref{sec:results}. Section \ref{sec:discussion} outlines the physical scenario and offers theoretical interpretations of these results. Finally, we present the conclusions in Section \ref{sec:conclusion}. Throughout this work, we adopt a distance of $d = 3.4$ kpc \citep{Mata2025A&A...693A.129M} and a black hole mass of $M_{\rm BH} = 10 \rm M_{\odot}$.

\section{Observations and data reduction} \label{sec:data}

In X-ray, Insight-HXMT performed long-term monitoring of Swift J1727.8-1613 at a cadence of less than 1 day. We use the Insight-HXMT Data Analysis software ({\tt HXMTDAS, v2.06}) to reduce the data, filtering the data with the default configuration\footnote{http://hxmtweb.ihep.ac.cn/SoftDoc/847.jhtml}. The joint spectra from the low-energy (LE), medium-energy (ME), and high-energy (HE) telescopes are extracted by screening the event files using different good time intervals (GTIs). For GTIs with short exposure, we combine nearby GTIs to improve the signal-to-noise ratio of the spectra. For simplicity, we selected the ME GTIs as the input for event screening, since the GTI time range in ME encompasses those in LE and HE for most observations. Then we obtain all spectra separated by the ME GTIs for subsequent analysis. 

\begin{figure*}
    \centering
    \includegraphics[width=0.9\textwidth]{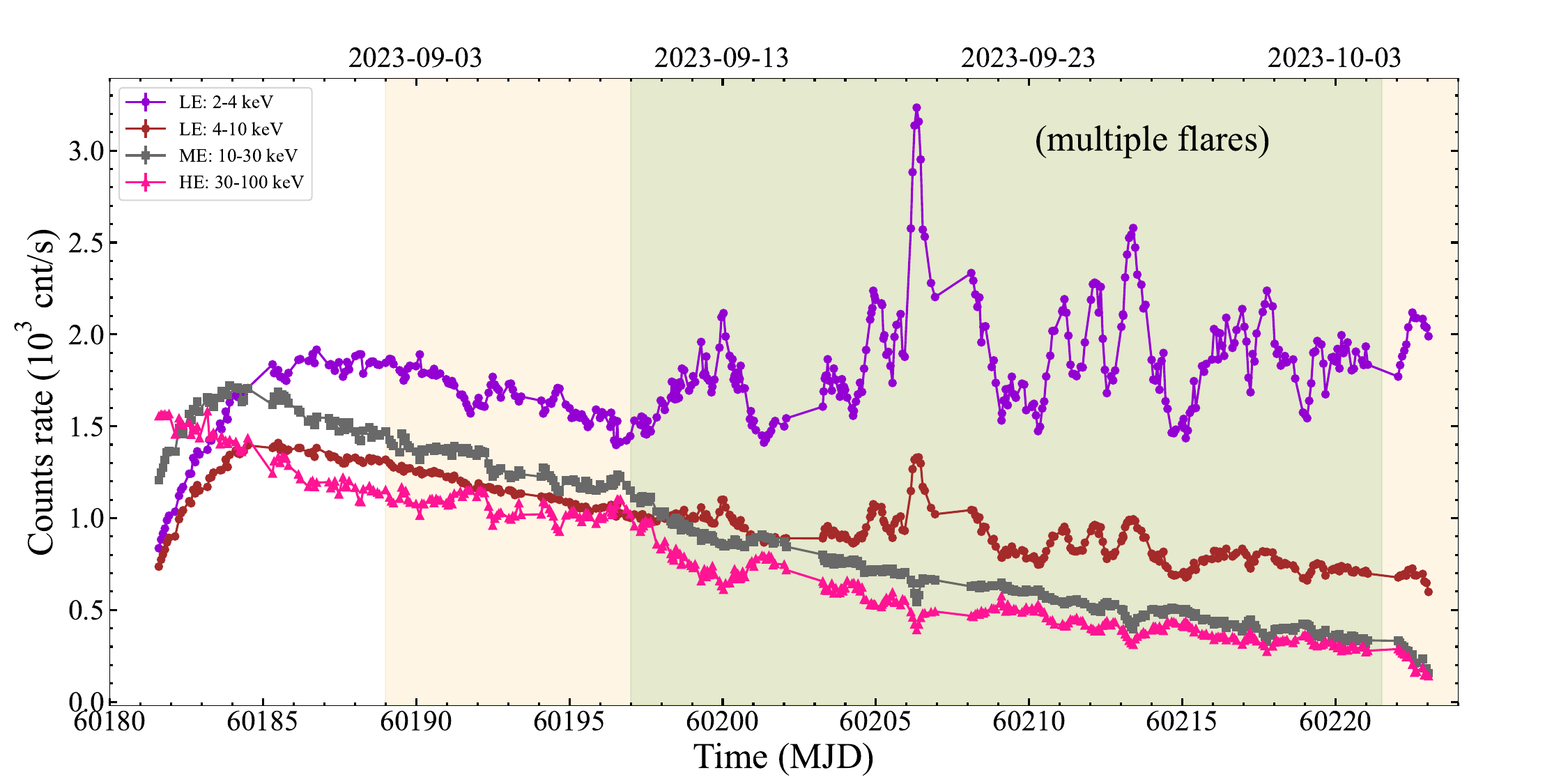}
    \caption{The X-ray light curve of Swift J1727.8-1613, as monitored by HXMT, is shown in the LE bands (2-4 keV in purple, 4-10 keV in brown), the ME band (10-30 keV in gray), and the HE band (30-100 keV in pink). The unshaded region and the orange region represent the hard state and the state transition, respectively. Multiple flares during the state transition are marked by the green region. Error bars are negligible and not visible for most data points. For visualization, the data in the ME and HE bands are scaled down by factors of 0.9 and 0.7, respectively.}
    \label{fig:lc}
\end{figure*}

The Insight-HXMT light curves, shown in Figure \ref{fig:lc}, span from 2023-08-25 (MJD 60181) to 2023-10-06 (MJD 60223) across three energy bands: 2-10 keV (LE), 10-30 keV (ME), and 30-100 keV (HE). The unshaded and orange regions represent the hard state and the state transition, respectively. The light curve shows that variations are significantly larger in the LE band than in the ME and HE bands. The LE count rate in the 2-4 keV range surged to approximately 1900 cts/s during the rising hard state on MJD 60186, then declined to around 1400 cts/s on MJD 60197. Following this, a variable LE light curve in the 2-4 keV band was observed, marked by a sequence of notable and intense flares from MJD 60197 to MJD 60223 (green region in Figure \ref{fig:lc}). The peak flux reached approximately 3300 cts/s during this interval. This suggests heightened activity in the accretion disk; however, spectral analysis is essential to identify the dominant component responsible for the low-energy photons.

The 4-10 keV light curve shows a trend similar to the 2-4 keV band but with slightly weaker variation, reaching a peak flux of approximately 1400 cts/s. The ME count rate rose from about 1300 cts/s on MJD 60181 to about 1900 cts/s on MJD 60184, then decreased to approximately 200 cts/s by MJD 60223. The HE count rate followed a similar pattern to the ME, though with a more gradual decline from about 2200 cts/s to about 220 cts/s throughout the outburst.

\section{Data analysis and results}
\label{sec:results}

\subsection{Basic Diagram}
\label{subsec:hid&rid}

\begin{figure}
    \centering
    \includegraphics[width=1\linewidth]{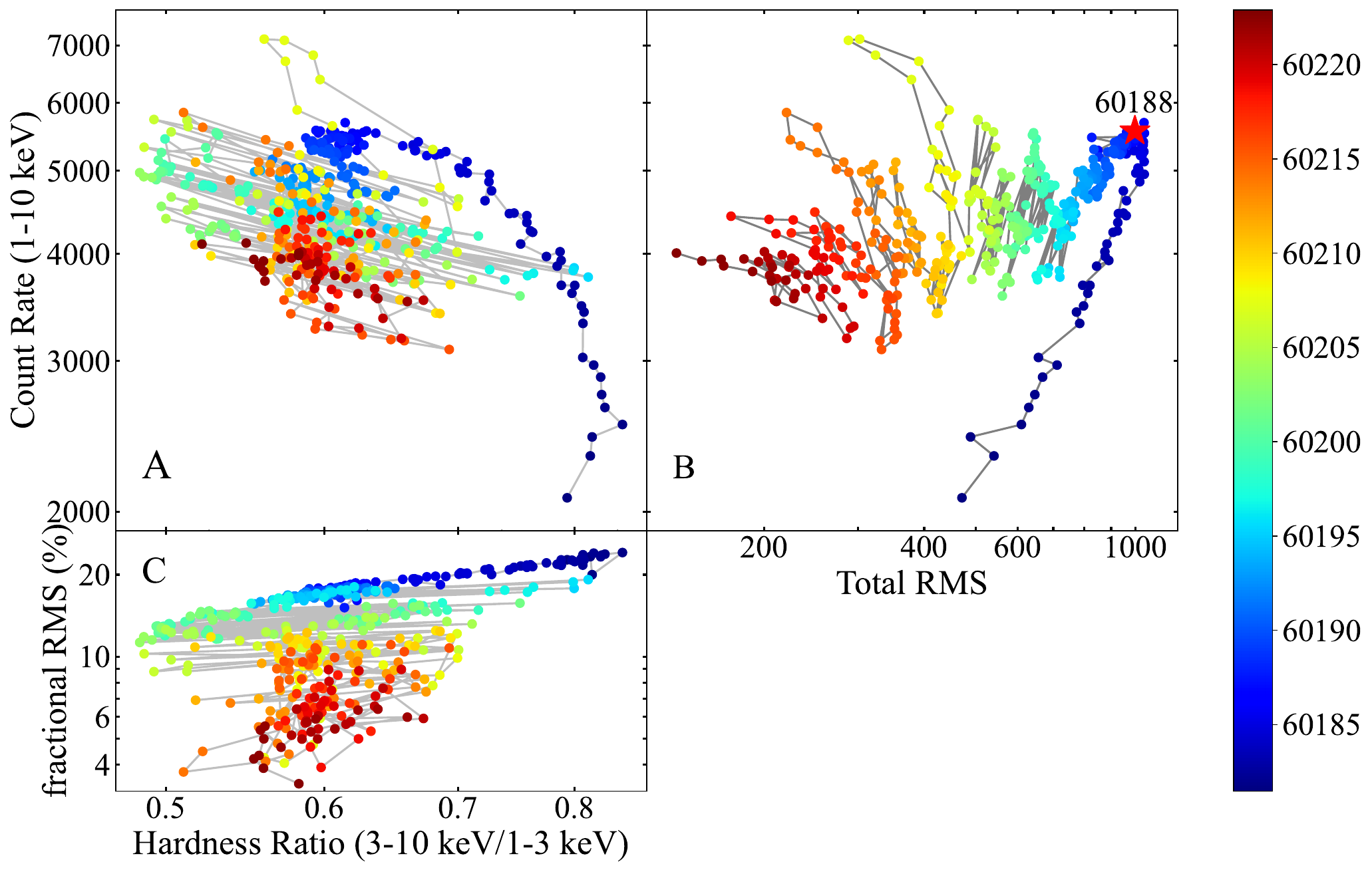}
\caption{Swift J1727.8-1613 outburst diagrams observed with HXMT: (A) Hardness-Intensity Diagram (HID); (B) RMS-Intensity Diagram (RID); (C) Hardness-RMS Diagram (HRD). The black star marks the onset of the hard-to-soft state transition. Error bars are negligible for most data points, and each point's color indicates the observation time.}
    \label{fig:basic diagram}
\end{figure}

The three diagrams, the hardness-intensity diagram (HID), the RMS-intensity diagram (RID), and the hardness-RMS diagram (HRD), are useful for investigating the evolution of the outburst and identifying the spectral states \citep{Belloni2016ASSL..440...61B}.
In the RID, the RMS is integrated over a broad frequency range in the power density spectra (PDS; \citealt{Belloni2016ASSL..440...61B}). We used the {\tt powspec} package in {HEAsoft} to compute the PDS for each HXMT detector observation, using a 64 s interval and a corresponding 1/128 s time resolution. After subtracting Poisson noise, the PDS is normalized using the Miyamoto normalization \citep{Miyamoto1991ApJ...383..784M}. The diagrams are plotted in Figure \ref{fig:basic diagram}. Here, the intensity is the count rate in the 1-10 keV band. The hardness ratio is defined as the ratio of the count rate in the 3-10 keV band to that in the 1-3 keV band. The RMS is integrated over the broad frequency range 0.1-60 Hz in the PDS. The source began the outburst at the bottom-right of the HID and then moved to the upper-left. In the RID, a diagonal line extending from the lower-left to the upper-right indicates that the source is in the hard state \citep{Belloni2016ASSL..440...61B}. Subsequently, deviations from the diagonal line indicate that the source entered the state transition around MJD 60188. These results are consistent with the criterion based on QPO evolution \citep{mata2024A&A...682L...1M}. 

\subsection{Spectral analysis}
\label{subsec:spectral analysis}

\begin{figure}
    \centering
    \includegraphics[width=1\linewidth]{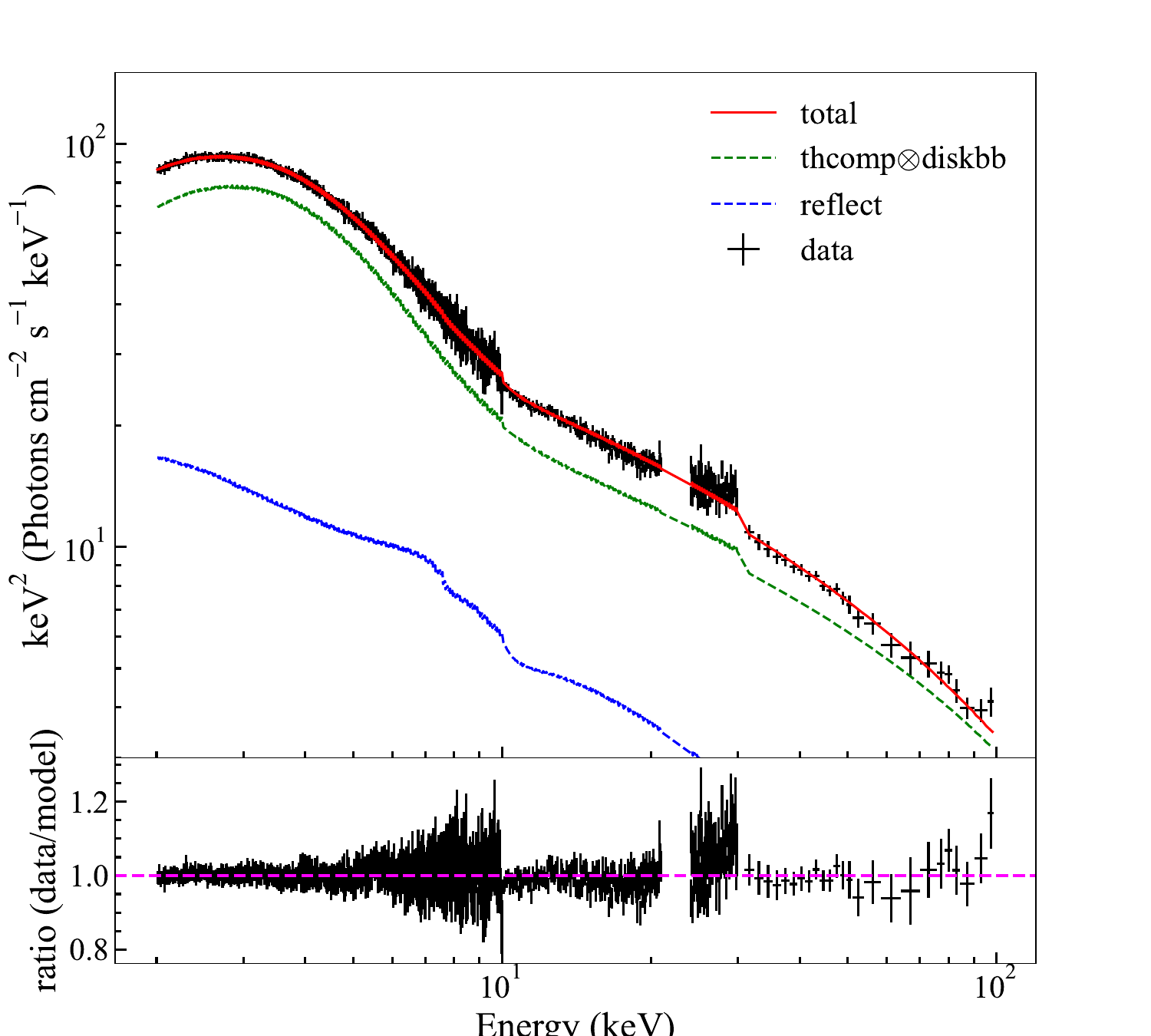}
    \caption{Upper panel: The example unfolded energy spectrum of Swift J1727.8-1613 at the source's brightest. The black points represent the observational data. The red line represents the total model. The green and blue dashed lines represent {{\tt thcomp$\otimes$diskbb}} and {\tt relxillCp}. Lower panel: The ratio of the data to the best-fitting model. 
    %The dashed purple line represents ratio=1 for reference. 
    Data at energies above 21-24 keV are excluded due to the fluorescence line. The spectrum's error bars correspond to the 1-$\sigma$ level. Note that data points are rebinned for visual clarity.}
    \label{fig:spec_fit}
\end{figure}

\begin{figure}
    \centering
    \includegraphics[width=0.5\textwidth]{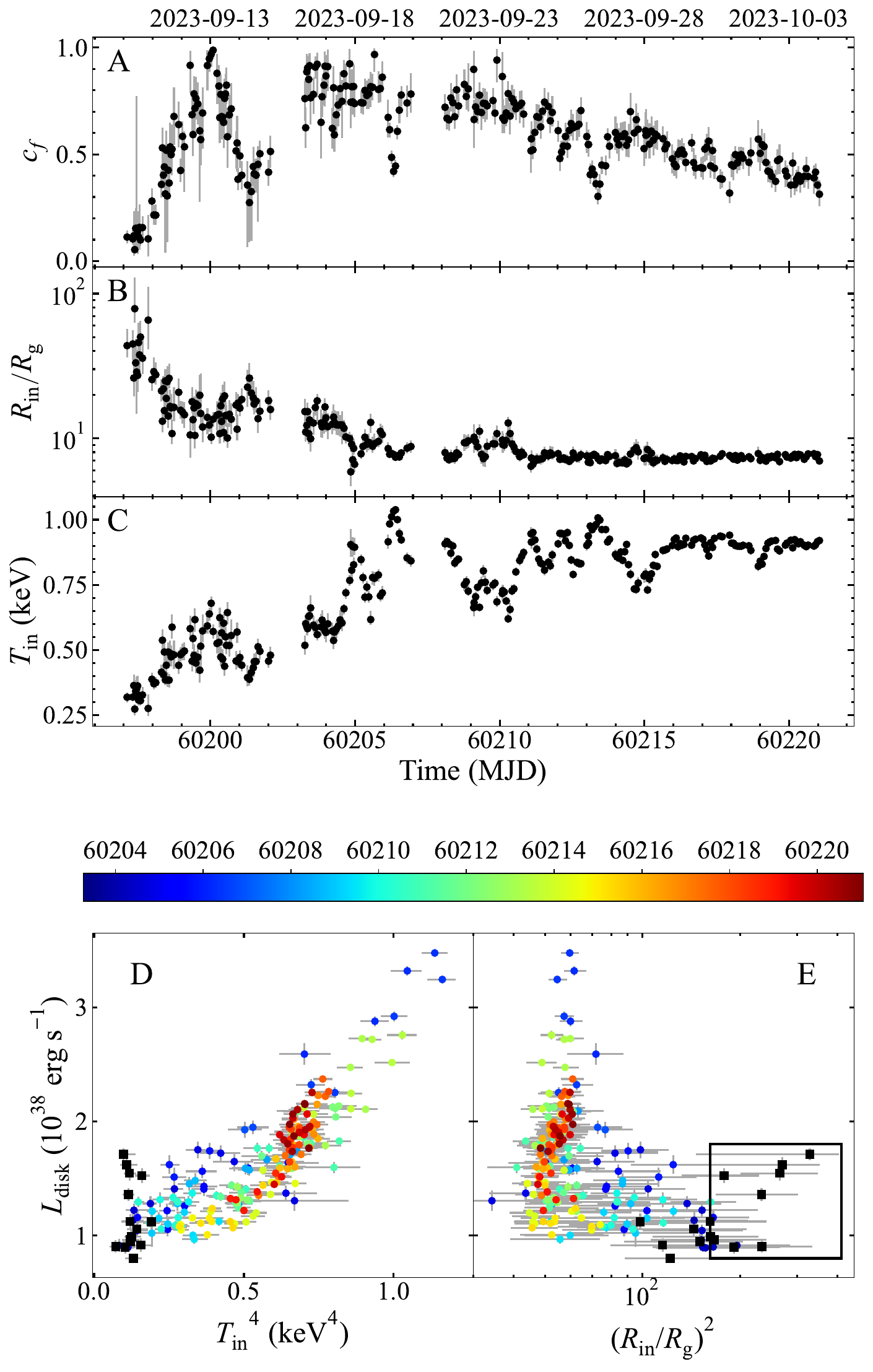}
    \caption{Panel A to C: Temporal evolution of parameters estimated from spectral fitting: A) the disk temperature at the inner radius, $T_{\rm in}$; B) the inner radius, $R_{\rm in}$, in units of $R_{\rm g}$; C) the covering fraction, $c_f$. Panels D to E: Correlations between disk luminosity $L_{\rm disk}$ and $T_{\rm in}^4$ and $R_{\rm in}^2$, respectively. The color of each point corresponds to the time of observation. The black squares represent data from MJD 60203-60204. The black rectangle highlights data for which $R_{\rm in}$ is poorly constrained. }
    \label{fig:spec_para}
\end{figure}

To identify the dominant component underlying the multiple flares observed during the state transition, we focus on the flaring phase between MJD 60197 and MJD 60223 (the green region in Figure \ref{fig:lc}) and analyze spectral data from HXMT observations. The 21-24 keV spectrum is excluded because of the photoelectric effect on electrons in the Silver K-shell of the ME detector. A systematic error of 1.5$\%$ is included in the spectral fitting \citep{You2021NatCo..12.1025Y}. We find that the spectrum is well fitted by the model {\tt constant*tbabs*(thcomp$\otimes$diskbb+relxillCp)} in {\tt XSPEC}. The {\tt tbabs} model \citep{Wilms2000ApJ...542..914W} represents Galactic absorption with a fixed column density of $N_{\rm H} = 0.226\times 10^{22}~\rm cm^{-2}$ \citep{OConnor2023ATel}. The {\tt thcomp} model describes spectra arising from Comptonization by hot thermal electrons \citep{Zdziarski2020MNRAS.492.5234Z}. The {\tt diskbb} model accounts for thermal emission from the disk. The {\tt relxillCp} model is a relativistic reflection model that includes both the direct emission from the corona and its reflection on the disk \citep{Garcia2014ApJ...782...76G}.  
We assume the iron abundance $A_{Fe}=1.0$ in solar units. The inclination prefers to be low value, and we fixed it at $\theta=40 ^{\circ}$ \citep{peng2024ApJ...960L..17P}. The reflection fraction $R_{\rm f}$ is defined in the rest frame as the ratio of intensity emitted towards the disk compared to escaping to infinity, and we set $R_{\rm f}=-1$ to obtain the reflected component \citep{Dauser2016A&A...590A..76D}. The outer radius of the reflection disk is fixed at the upper limit of their table model $R_{\rm out}=1000R_{\rm g}$, where $R_{\rm g}=GM/c^2$ is the gravitational radius.
The input inner radius in {\tt relxillCp} model is linked to the normalization in {\tt diskbb}, which directly relates to the true inner disk radius \citep{Kubota1998PASJ...50..667K,Basak2016MNRAS.458.2199B}:
\begin{equation}
    N_{\rm diskbb} = \frac{(R'_{\rm in}/1~{\rm km})^2\cos{\theta}}{(d/10~{\rm kpc})^2}.
    \label{eq:rin}
\end{equation}
Here, $R'_{\rm in}$ is the apparent inner radius, $\theta$ is the inclination, and $d$ is the distance. The intrinsic inner radius $R_{\rm in}$ relates to the observed apparent radius $R'_{\rm in}$ as $R_{\rm in} = \kappa^2 \zeta R'_{\rm in}$, with the color correction factor $\kappa=1.7$ for a diluted blackbody \citep{Shimura1995ApJ...445..780S}. The parameter $\zeta$ accounts for the inner boundary condition correction. According to \cite{Kubota1998PASJ...50..667K}, $\zeta\simeq0.41$ applies to a non-spinning black hole with a disk extending to the innermost stable circular orbit (ISCO). Although the black hole in Swift J1727.8-1613 likely has a spin of $a>0$ \citep{peng2024ApJ...960L..17P, Yu2024MNRAS.529.4624Y}, it remains unclear whether the inner disk reaches the ISCO during the state transition. Therefore, we adopt $\zeta = 1$ to neglect this correction in this work. As an example, the unfolded X-ray spectrum and the best-fit model when the source is brightest are plotted in Figure \ref{fig:spec_fit}. The temporal evolution of key parameters, such as the covering fraction $c_f$, inner disk radius $R_{\rm in}$, and the corresponding temperature $T_{\rm in}$, is presented in Figure \ref{fig:spec_para}.

We then use the task {\tt cflux} in {\tt XSPEC} to estimate the intrinsic disk flux $F_{\rm disk}$ and the X-ray Comptonization flux $F_{\rm Comp}$ in the 0.01-1000 keV range. $F_{\rm disk}$, composed of the seed photons for Compton cooling and the unscattered soft photons, is directly estimated using the {\tt diskbb} model. The flux estimated by {\tt thcomp} includes the Comptonized photons and the unscattered soft photons from the disk \citep{Zdziarski2020MNRAS.492.5234Z}. Therefore, given the covering fraction $c_f$---the fraction of disk low-energy photons intercepted by the hot inner flow---the seed-photon flux $F_{\rm seed}$ and the Comptonization flux $F_{\rm Comp}$ can be estimated as $F_{\rm seed}=c_fF_{\rm disk}$ and $F_{\rm Comp}=F_{\rm thcomp}-(F_{\rm disk} - F_{\rm seed})=F_{\rm thcomp}-(1-c_f)F_{\rm disk}$, respectively \citep{Zdziarski2020MNRAS.492.5234Z}. The disk and Compton luminosities are then calculated at a distance of $d=3.4\,\text{kpc}$.

The Compton luminosity does not show significant multiple flares, as in the disk emission, but includes a transient flare period during MJD 60197-60202, resembling the LE light curve. This behavior is caused by the covering fraction hump in this phase, which leads to varying seed photons for Compton cooling. Following this, the Compton luminosity rose until MJD 60206 before it gradually decreased.

The disk component exhibits significant variability, appearing as a sequence of damped multi-flares superimposed on a slight upward trend. The disk luminosity was initially low before MJD 60203, producing only a minor change in flux. Afterward, the disk emission shows significant X-ray flux modulation throughout the transition, with the temporal evolution marked by reduced amplitude and a shortened modulation period, resembling the LE light curve. The most intense disk flare around MJD 60205 boosted the luminosity to $3.5 \times 10^{38} \, \text{erg} \, \text{s}^{-1}$, a $\sim$3.9-fold increase, and progressively damped to a lower peak of $2.3 \times 10^{38} \, \text{erg} \, \text{s}^{-1}$ around MJD 60221, corresponding to a $\sim$1.8-fold rise. Such a decrease in peak luminosity and reduced variability suggest damping of disk emission fluctuations, mirroring the trend in the LE light curve.

%%%%%%%%%%%%%%%%%%%%spectral analysis%%%%%%%%%%%%%%%%%%%%%%

\begin{figure*}
    \centering
    \includegraphics[width=0.9\textwidth]{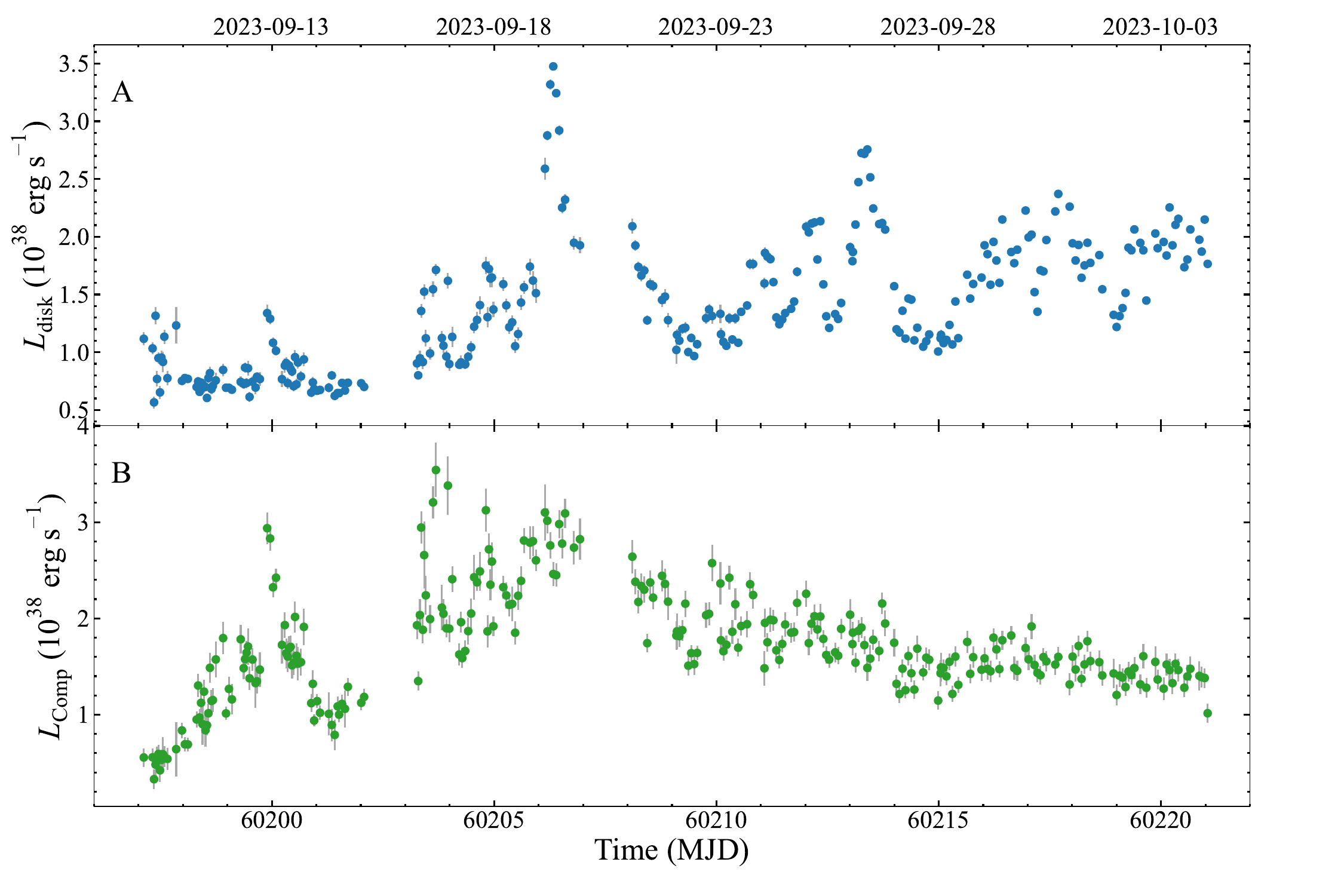}
    \caption{The luminosity evolution for the spectral components in 0.01-1000 keV is shown. Panel A presents the disk luminosity $L_{\rm disk}$
    %consisting of the intercepted $L_{\rm seed}$ and observed photons
    , whereas Panel B illustrates the Compton luminosity $L_{\rm Comp}$.}
    \label{fig:lum}
\end{figure*}

\section{discussion}
\label{sec:discussion}

Before transitioning to a soft state, the recently identified Swift J1727.8-1613 remained in the state transition for approximately 43 days. The HXMT captured this state transition at a high cadence. Intriguingly, the LE instrument recorded a significant variation in count rates from 2-10 keV (especially in the 2-4 keV band), whereas the ME and HE instruments observed a moderate decrease in count rates above 10 keV. Spectral analysis revealed, for the first time, that the disk exhibits pronounced daily variability and multi-flare behavior, while the Compton luminosity shows a slight decrease during the state transition. 

%============================================
\subsection{Origin of the disk flares}
\label{subsec:scenario}

According to the description of {\tt diskbb}, the disk emission can be estimated using the simplified scaling relation $ L_{\rm disk} \propto T_{\rm in}^4 R_{\rm in}^2 $, where $R_{\rm in}$ is the inner disk radius and $ T_{\rm in} $ is the corresponding temperature at this boundary. In this work, the temperature $T_{\rm in}$ is measured directly from the spectral analysis, while the inner disk radius $R_{\rm in}$ is estimated from the normalization in {\tt diskbb} (see Equation \ref{eq:rin}). The temporal evolutions of $T_{\rm in}$ and $R_{\rm in}$ are shown in Figures \ref{fig:spec_para} B and C. We find that the disk temperature is predominantly high, ranging from 0.6-1.0 keV, with damped oscillations centered around $ \sim 0.9 $ keV.

Concurrently, the inner radius $ R_{\rm in} $ exhibits small fluctuations before stabilizing near 8 $ R_{\rm g} $, indicating a thermal disk extending into the strong-gravity regime within a few gravitational radii. Thus, the observed damped variations in disk emission are hypothesized to arise from oscillations in $ T_{\rm in} $ rather than from a quasi-stable $R_{\rm in}$. This interpretation is strongly supported by the distinct temperature dependence of $L_{\rm disk}$, as shown in Figure \ref{fig:spec_para}D. We find a robust correlation with Spearman's coefficient $r_{\rm S}=0.87$ (significance $>99.9\%$), whereas the $L_{\rm disk}-R_{\rm in}^2$ relation remains statistically marginal ($r_{\rm S}=-0.32$ at the 4.7$\sigma$ level, see Figure \ref{fig:spec_para}E). These results unambiguously identify temperature modulation as the principal driver of disk emission variability. Furthermore, since {\tt diskbb} neglects the inner boundary, $T_{\rm in}$ can be expressed as a function of the mass accretion rate $\dot{M}_{\rm in}$ and the inner disk radius $R_{\rm in}$, following $T_{\rm in}\propto (\dot{M}_{\rm in}/R_{\rm in}^3)^{1/4}$. Given the quasi-stable $R_{\rm in}$, it can be inferred that the accretion rate dominates the temperature variations and thereby generates the fluctuation in the disk emission. However, $R_{\rm in}$ exhibits exceptionally large uncertainties (see black rectangle in Figure \ref{fig:spec_para}E) for most observations during MJD 60203–60204, making the temporal evolution of the accretion rate statistically unconstrained. Consequently, data during this period have been excluded from the subsequent analysis, as severe parameter degeneracies between $R_{\rm in}$ and $\dot{M}$ preclude reliable physical interpretation of the observed variations.

%==================================================
\subsection{The $\Gamma$-flux relation} \label{subsec:gamma-flux}

Having established that the disk fluctuations are driven by variations in the inner accretion rate, we now investigate the joint evolution of the disk and corona throughout the transition, as revealed by the spectral parameters. It is well established that the photon index $\Gamma$ and X-ray luminosity $L_{\rm X}$ are correlated \citep{yuan2007ApJ...658..282Y,wu2008ApJ...682..212W,caoxf2014ApJ...788...52C,yangqx2015MNRAS.447.1692Y}. Below 1-2\% of the Eddington luminosity $L_{\rm Edd}$, $L_{\rm X}$ shows a negative correlation with $\Gamma$. In contrast, luminous sources exhibit a positive correlation between $\Gamma$ and $L_{\rm X}$.

\begin{figure*}
    \centering
    \includegraphics[width=1\linewidth]{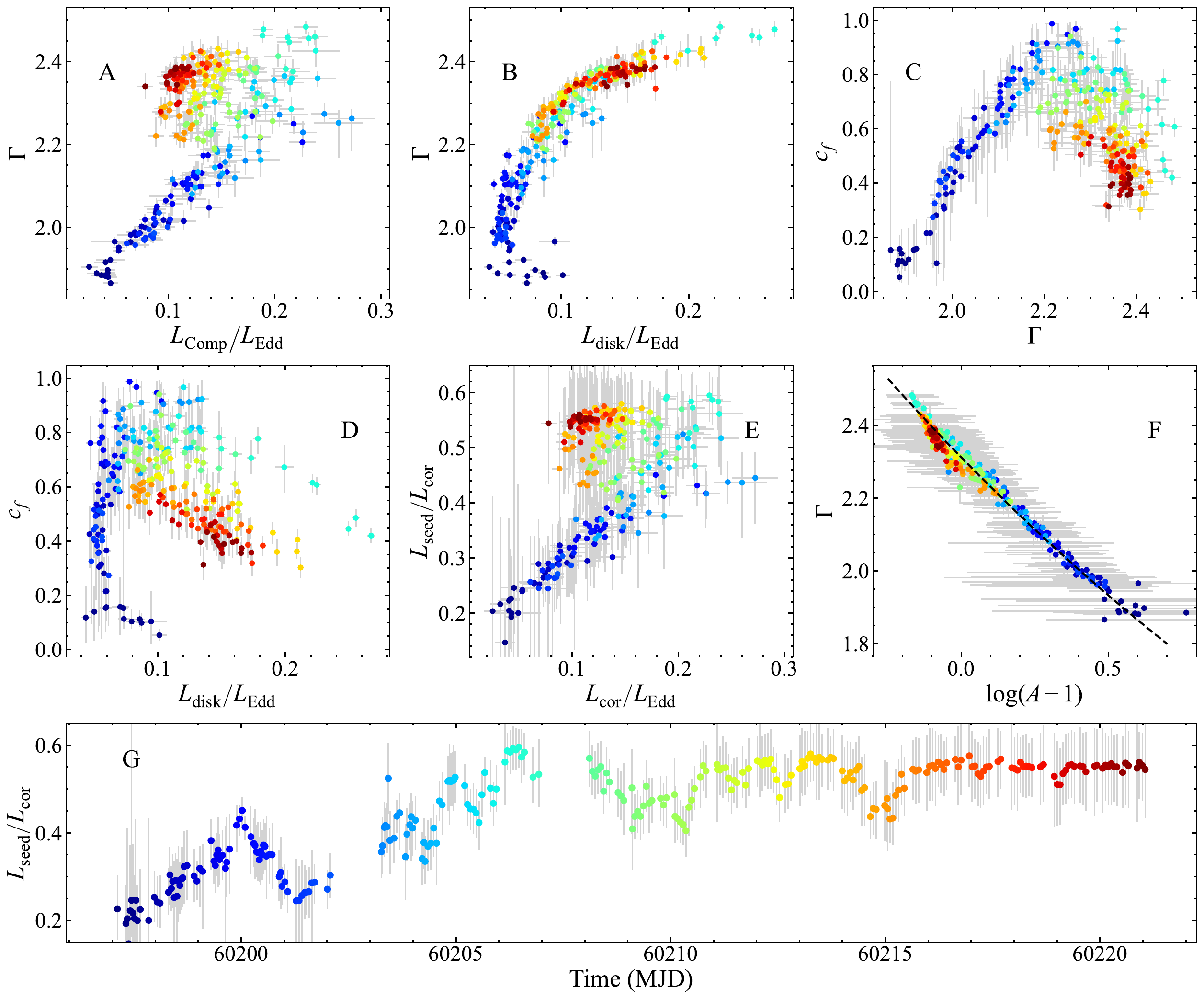}
    \caption{A: The relationship between the photon index $\Gamma$ and the Eddington-scaled Compton luminosity; B: The relationship between the photon index $\Gamma$ and the Eddington-scaled disk luminosity; C: The correlation between $c_f$ and $\Gamma$; D: The correlation between $c_f$ and $L_{\rm disk}/L_{\rm Edd}$; E: The correlation between $L_{\rm seed}/L_{\rm cor}$ and the Eddington-scaled Compton luminosity $L_{\rm Comp}/L_{\rm Edd}$. F: The relation between the photon index $\Gamma$ and the Compton amplification factor. The dashed line represents the formula in \cite{Beloborodov1999ASPC..161..295B}; G: The temporal evolution of $L_{\rm seed}/L_{\rm cor}$. The color of each point corresponds to the time of observation.}
    \label{fig:gamma_flux}
\end{figure*}

Figure \ref{fig:gamma_flux}A shows the correlation between the photon index $\Gamma$ and the Compton luminosity $L_{\rm Comp}$, scaled to the Eddington luminosity, across all epochs. The spectral slope is harder during the first flare phase before MJD 60203, with $\Gamma<2.15$ for most observations, whereas it is systematically softer ($\Gamma \geq 2.15$) during the subsequent flares. Notably, $\Gamma$ is positively correlated with $L_{\rm Comp}$ during each flare, consistent with previous findings in this high-luminosity range. After MJD 60203, Figure \ref{fig:gamma_flux}A shows a dynamic correlation between the photon index $\Gamma$ and $L_{\rm Comp}$. Specifically, the positive correlation shifts toward the upper-left corner of the plane, characterized by a reduced range in both $\Gamma$ and $L_{\rm Comp}$. The relation between the Eddington-scaled disk luminosity $L_{\rm disk}$ and $\Gamma$ is also plotted in Figure \ref{fig:gamma_flux}B, showing a significant positive relationship with a Spearman's coefficient of $r_{\rm S} = 0.94$ at a confidence level of $>99.9\%$.

Previous work attributed this positive correlation to the faster increase in the power of soft seed photons $L_{\rm seed}$ relative to the power supplied to the electrons in the hot accretion flow $L_{\rm cor}$ \citep{Zdziarski2002ApJ...578..357Z, yangqx2015MNRAS.447.1692Y}. Specifically, when $L_{\rm cor}$ is fixed, an increase in $L_{\rm seed}$ enhances Compton cooling, resulting in a softer energy spectrum \citep{Zdziarski2002ApJ...578..357Z}. Consequently, the photon index $\Gamma$ positively correlates with the power ratio $L_{\rm seed}/L_{\rm cor}$. In the case of soft seed photons originating from the accretion disk around a BH, $\Gamma$ can be estimated using \citep{Beloborodov1999ASPC..161..295B},
\begin{equation}
    \Gamma = \frac{7}{3}(A-1)^{-\frac{1}{6}},
    \label{eq:gamma}
\end{equation}
where $A = L_{\rm cor}/L_{\rm seed}$ is the Compton amplification factor. Thus, the $\Gamma-L_{\rm Comp}$ correlation can be understood by analyzing the relationship between $ L_{\rm seed}/L_{\rm cor}$ and $L_{\rm cor}$. In the following section, we apply this framework to interpret the observed dynamic $\Gamma-L_{\rm Comp}$ correlation shown in Figure \ref{fig:gamma_flux}.

First, it is essential to estimate both $L_{\rm cor}$ and $L_{\rm seed}$. Given the high Eddington ratio $L_{\rm Comp}/L_{\rm Edd}$ in this study (Figure \ref{fig:gamma_flux}A), nearly all the energy absorbed by electrons in the hot accretion flow is efficiently radiated away \citep{Xie2012MNRAS.427.1580X}. This supports estimating the Compton luminosity $L_{\rm Comp}$ as $L_{\rm cor}$ \citep{yangqx2015MNRAS.447.1692Y}, i.e., $L_{\rm Comp} \approx L_{\rm cor}$. Additionally, in the bright X-ray luminosity regime, it is hypothesized that the soft seed photons for Compton cooling originate from the accretion disk \citep{yangqx2015MNRAS.447.1692Y}. However, equating $L_{\rm disk}$ directly with $L_{\rm seed}$ would be inaccurate because not all soft photons are scattered due to the geometrical effects of the hot accretion flow. Through spectral fitting, we derive the evolution of the covering fraction $c_f$. Consequently, if the contribution from self-absorbed synchrotron emission in the hot flow is disregarded, the soft seed photons for Compton cooling can be approximated as $L_{\rm seed} = L_{\rm disk} \cdot c_f$.

Figure \ref{fig:gamma_flux}C and \ref{fig:gamma_flux}D present the correlations between $c_f$ and $\Gamma$ and $L_{\rm disk}$. Prior to MJD 60203, $c_f$ exhibits a strong positive correlation with $\Gamma$, with a Spearman coefficient $r_{\rm S}=0.97$ at a significance of $>99.9\%$. In contrast, no significant correlation is found between $c_f$ and $L_{\rm disk}$ ($r_{\rm S}=0.10$, 0.86$\sigma$ confidence). After MJD 60203, both correlations reverse, as $c_f$ becomes moderately and negatively correlated with $\Gamma$ ($r_{\rm S}=-0.56$, $>99.9\%$ confidence) and with $L_{\rm disk}$ ($r_{\rm S}=-0.59$, $>99.9\%$ confidence). It is inferred that the increase in $\Gamma$ at low $L_{\rm disk}$ is driven almost entirely by increasing $c_f$ at a nearly constant $L_{\rm disk}$. For higher $L_{\rm disk}$, while $c_f$ decreases, $L_{\rm disk}$ increases even more significantly, leading to a net softening of the spectrum. Figure \ref{fig:gamma_flux}E shows the correlation between $L_{\rm seed}/L_{\rm cor}$ and $L_{\rm Comp}/L_{\rm Edd}$. A clear positive correlation is observed for each flare, indicating that $L_{\rm seed}$ increases more rapidly than $L_{\rm cor}$. According to the analysis in \cite{Zdziarski2002ApJ...578..357Z}, $\Gamma$ is expected to correlate positively with $L_{\rm Comp}$. We also examine the relationship between $\Gamma$ and $A$, as shown in Figure \ref{fig:gamma_flux}F. The spectral results are in excellent agreement with the theoretical framework proposed in \cite{Beloborodov1999ASPC..161..295B} within uncertainties.

Furthermore, Figure \ref{fig:gamma_flux}A shows a dynamic correlation in which positive correlations shift toward the upper-left of the plane, marked by a reduced range in $\Gamma$. According to Eq. \ref{eq:gamma}, variability in $L_{\rm disk}$ (i.e., $L_{\rm seed}$), combined with the mild decrease in $L_{\rm Comp}$, drives changes in $L_{\rm seed}/L_{\rm cor}$ (see Figure \ref{fig:gamma_flux}G), producing the observed dynamic $\Gamma-L_{\rm Comp}$ relationship. Additionally, the progressive damping of $L_{\rm disk}$ fluctuations weakens modulation of $L_{\rm seed}$, thereby reducing the range of $L_{\rm seed}/L_{\rm cor}$ and suggesting a narrowing of $\Gamma$. Concurrently, the global increase in $L_{\rm seed}/L_{\rm cor}$ indicates a higher $\Gamma$. As $L_{\rm Comp}$ decreases, the $\Gamma-L_{\rm Comp}$ relation shifts toward the upper-left of the plane, accompanied by narrowing ranges in both $\Gamma$ and $L_{\rm Comp}$.

\subsection{Viscous damping in the accretion disk}
\label{subsec:propagating model}

In Section \ref{subsec:scenario}, we demonstrated that daily damped fluctuations in disk emission are primarily driven by variations in the inner accretion rate. However, a thermal disk extending close to ISCO rules out local generation of such fluctuations in the inner disk, as the associated timescales (typically tens of seconds for a stellar-mass black hole, see \citealt{King2004MNRAS.348..111K, Uttley2005MNRAS.359..345U, Ingram2012MNRAS.419.2369I, Mushtukov2018MNRAS.474.2259M, Zhan2025arXiv250203995Z}) are orders of magnitude shorter than the observed daily variability. We therefore propose that the observed disk emission variations likely originate from accretion rate fluctuations generated at large radii, with progressively shortened periods, that then propagate inward \citep{Lynden-Bell1974MNRAS.168..603L, Lyubarskii1997MNRAS.292..679L, Kotov2001MNRAS.327..799K}. As the period shortens, viscous damping of the propagating fluctuation strengthens, naturally reducing the amplitude of the accretion rate variations at the inner disk and hence the observed disk emission fluctuations (see Appendix \ref{appendix} and \citealt{Zdziarski2009MNRAS.399.1633Z}). 

\begin{figure*}
    \centering
    \includegraphics[width=0.8\textwidth]{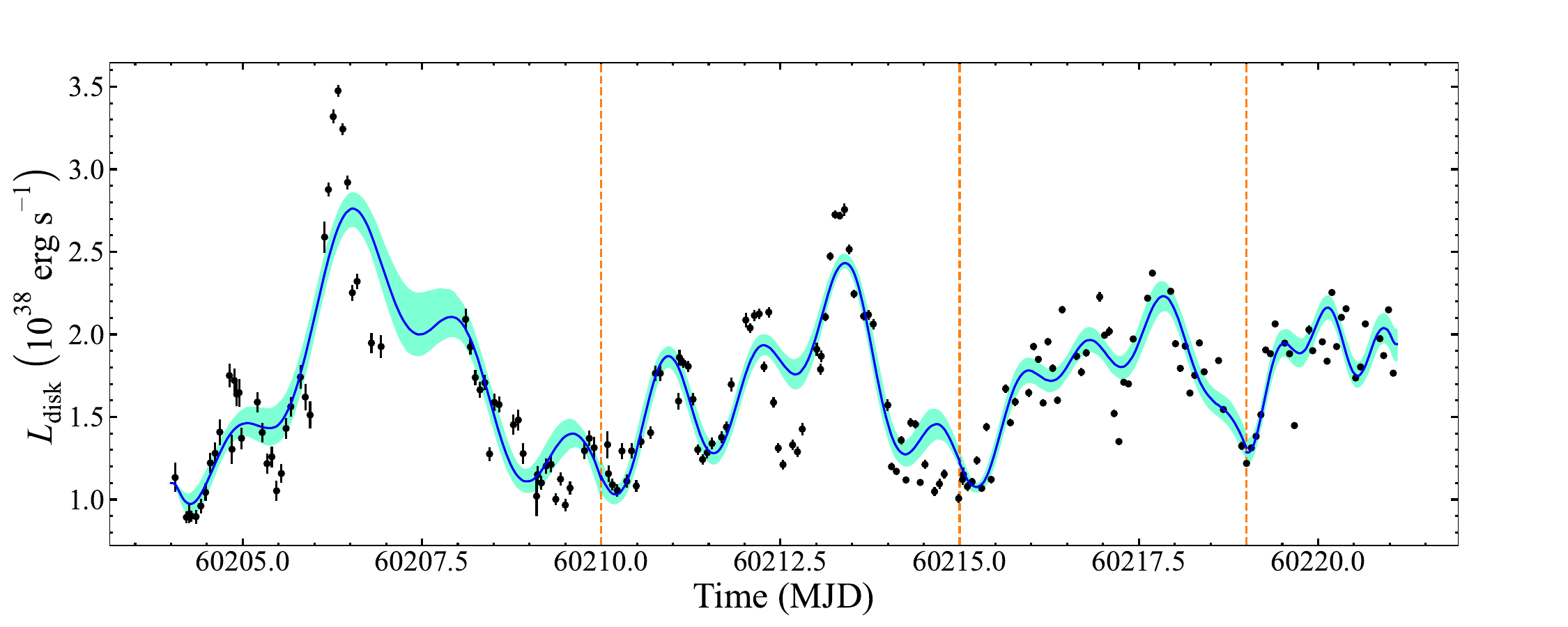}
    \caption{Fits to the disk flares using the propagating fluctuation model, assuming the disk's inner radius is constant. The black dots represent the disk luminosity, while the blue line shows the best fit from the MCMC approach. The green shaded region indicates the 1-$\sigma$ uncertainties. The orange dashed lines mark the division lines for the four epochs (see Appendix \ref{appendix}).}
    \label{fig:damping_fit}
\end{figure*}

During disk flares, the inner disk radius remains quasi-stable, allowing us to approximate $L_{\rm disk} = \eta \dot{M}_{\rm in}c^2$ with a constant $\eta$ and to analyze the disk emission directly rather than the accretion rate. We use the Markov chain Monte Carlo (MCMC) approach implemented in the {\tt emcee} package in {\tt Python} \citep{Foreman_Mackey_2013} to fit disk variability within the framework of the propagating-fluctuation model. The modeling method is detailed in Appendix \ref{appendix}. We emphasize that the purpose of this modeling is not to reproduce every detail of the disk light curve or to derive precise values of physical parameters (such as the viscous timescale). Rather, we aim to demonstrate that the observed damping of flare amplitude arises naturally from the progressive shortening of fluctuation periods as they propagate inward, which is qualitatively consistent with expectations of the propagating-fluctuation model \citep{Zdziarski2009MNRAS.399.1633Z}. The fitting results are presented in Figure \ref{fig:damping_fit}. 

Although the light curve was divided into four epochs (each spanning roughly one cycle of the fundamental period) and the outer fluctuations were approximated with harmonically related cosine functions (an idealization that cannot fully capture the stochastic nature of the variability), the model successfully reproduces the key morphological features across all epochs, including systematic amplitude damping and flare periods shortening. Importantly, the fitted amplitudes of the outer fluctuations source remain roughly constant across epochs, indicating that the observed amplitude decrease is predominantly caused by enhanced viscous damping during inward propagation rather than by intrinsic weakening of the external driver. Our main conclusions are robust to reasonable shifts in the epoch boundaries of $\pm$ 1 day.

\begin{figure}
    \centering
    \includegraphics[width=1\linewidth]{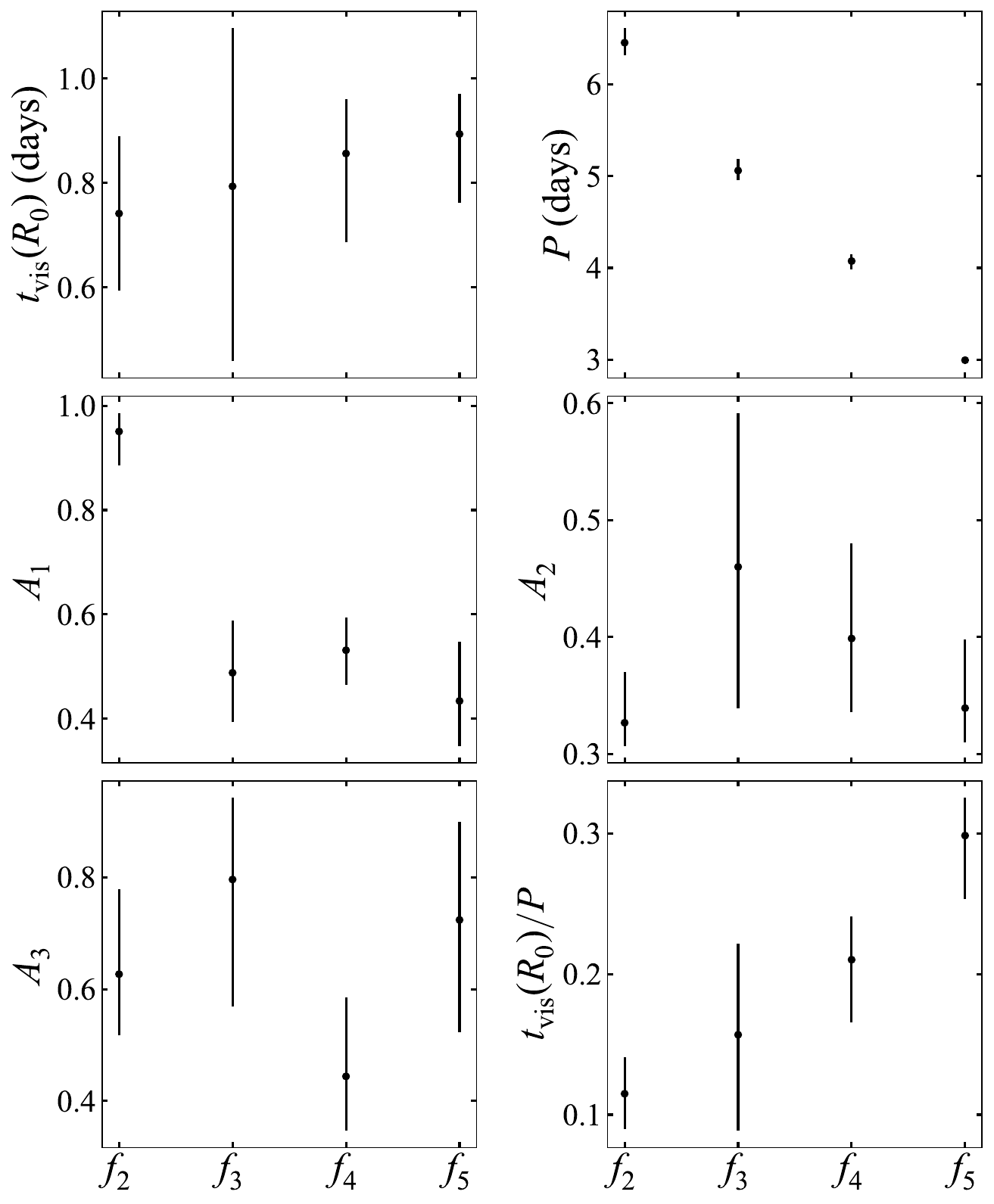}
    \caption{The evolution of the main parameters in the propagating fluctuation model. The dots represent the median values for each parameter. The error bars are quoted at the 1-$\sigma$ level. Note that $f_2$ to $f_5$ are linked to the four epochs in Figure \ref{fig:damping_fit}.}
    \label{fig:parameters}
\end{figure}

The key fitting parameters are shown in Figure \ref{fig:parameters}. We find that $t_{\rm vis}(R_0)$ remains approximately constant across epochs, while the fluctuation period decreases from 6.42 days to 2.99 days, driving a systematic increase in $t_{\rm vis}(R_0)/P$. As predicted by the propagating fluctuation model (Figure \ref{fig:greenfun}B; \citealt{Zdziarski2009MNRAS.399.1633Z}), this leads to stronger viscous suppression of the amplitude at $R_{\rm in}$, naturally explaining the observed amplitude damping. The fitted fluctuation source amplitudes $A_i$ remain nearly constant across epochs, suggesting that viscous diffusion is the dominant cause of the observed damping. The large $A_1$ at the earliest epoch may indicate intrinsic source weakening in the early phase. In addition, we note that $t_{\rm vis}$ and $A_i$ exhibit partial degeneracy (see the contour in Appendix \ref{appendix}). However, this degeneracy does not affect our main conclusion: the systematic increase in $t_{\rm vis}(R_0)/P$ across epochs is driven primarily by the progressive shortening of the period $P$, as evident from the median posterior values.

\subsubsection{A possible mechanism for the disk flares}

The next issue is how these daily fluctuations are generated in the outer disk.
Magneto-hydrodynamic (MHD) turbulence, driven the magnetorotational instability \citep{Balbus1991ApJ...376..214B,Balbus1998RvMP...70....1B}, is a primary mechanism for generating the fluctuations in the accretion disk, but operates on millisecond timescales in the inner disk region \citep{Lyubarskii1997MNRAS.292..679L, Kotov2001MNRAS.327..799K,Done2007A&ARv..15....1D,Ingram2012MNRAS.419.2369I, Mushtukov2018MNRAS.474.2259M,Zhan2025arXiv250203995Z}, i.e., much shorter than the daily variability observed here. 

One candidate is the thermal-viscous disk instability model (DIM; \citealt{Hameury1998MNRAS.298.1048H,Lasota2001NewAR..45..449L}), which predicts ``reflares'' on daily timescales consistent with Swift J1727.8-1613. In DIM, once the temperature and surface density exceed a critical threshold, a thermal instability launches an outward-propagating heating front. When the post-front density drops below the minimum on the hot branch, a cooling front initiates and propagates inward. The interplay between these fronts governs the flare morphology \citep{Lasota2001NewAR..45..449L,Dubus2019A&A...632A..40D}. As the disk density declines through accretion, the front propagation distance shortens, accelerating the heating–cooling cycle and producing a sequence of reflares with progressively shortened periods \citep{Dubus2001A&A...373..251D,Dubus2019A&A...632A..40D}. However, additional effects such as irradiation or disk winds can suppress reflares by inhibiting front propagation \citep{Dubus2001A&A...373..251D,Dubus2019A&A...632A..40D}.

The propagation speed of these fronts depends on the viscosity parameter $\alpha$ and the sound speed $c_S$ \citep{Meyer1984A&A...131..303M, Vishniac1996ApJ...471..921V, Lasota2001NewAR..45..449L}, which is
\begin{equation}
    V_{\rm front}\approx\alpha c_S.
\end{equation}
Given the radial extent of the moving front, the characteristic duration of a flare can be derived by
\begin{equation}
    t_{\rm front}=\frac{\Delta R}{V_{\rm front}}\approx\frac{\Delta R}{\alpha c_S}.
    \label{eq:t_front}
\end{equation}
The simulation of a stellar-mass BH suggested that the heating and cooling fronts propagate between $4\times10^{10}$ and $1\times10^{11}\,\text{cm}$ \citep{Dubus2019A&A...632A..40D}. Substituting $\Delta R=6\times10^{10}\,\text{cm}$ into Equation \ref{eq:t_front}, the characteristic front propagation timescale, which corresponds to half the fluctuation period $t_{\rm fluctuation}$, exhibits a linear dependence on black hole mass given by:
\begin{equation}
    t_{\rm fluctuation}=\frac{1}{2}t_{\rm front} \approx \frac{\Delta R}{V_{\rm front}}=\frac{\Delta R}{\alpha c_S}=3.45\frac{0.2}{\alpha}\,\text{days}.
\end{equation}
Here, $\alpha=0.2-1$ \citep{Tetarenko2018Natur.554...69T} and $c_S=10~\rm km/s$ are assumed (the front temperature is $\sim 10^4\rm~K$ due to hydrogen ionization; \citealt{Dubus2001A&A...373..251D,Lasota2001NewAR..45..449L}). We then derive a fluctuation period of approximately 6.9 days for $\alpha=0.2$.

%===============================================

\section{Conclusions}
\label{sec:conclusion}

In this study, we investigate the 2023 outburst of the newly discovered low-mass black hole X-ray binary Swift J1727.8-1613. High-cadence HXMT observations during its long transition revealed intense daily variations in the soft X-ray band, alongside a gradual decrease with modest variations in the high-energy band. The key findings of this study are summarized below:

\begin{enumerate}

    \item 
    The disk emission showed significant X-ray flux modulation throughout the transition, with temporal evolution marked by a dampened amplitude and a shortened modulation period. The modulation of disk emission is predominantly driven by fluctuations in the inner accretion rate rather than by geometric effects.

    \item 
     There is a positive correlation between $\Gamma$ and $L_{\rm Comp}$ during the transition. Moreover, we found that such correlations after MJD 60203 show temporal evolution, with the $\Gamma$-$L_{\rm Comp}$ correlation shifting toward the top-left of the plane, narrowing the range in $L_{\rm Comp}$ and increasing $\Gamma$. This behavior is intrinsically caused by fluctuations in the flux of the seed soft photons due to variable disk emission. 

    \item 
    We propose that the observed damped disk flares are due to accretion-rate fluctuations that are viscously damped as they propagate through the disk. This is further supported by directly fitting the observed disk flares within the propagating fluctuation model. Additional analysis suggests that the fluctuations originate from the reflare predicted by the accretion disk thermal instability model.

\end{enumerate}

\section*{Acknowledgements}

We thank Guillaume Dubus, Bi-Fang Liu, Xinwu Cao, Mouyuan Sun, Yue Wu, Di-Zhan Du, and Yi-Long Wang for their helpful discussions and comments.
This work is supported by NSFC grants 12322307, 12273026, 12361131579, and 12373049; supported by ``the Fundamental Research Funds for the Central Universities"; Xiaomi Foundation / Xiaomi Young Talents Program. The data analysis in this paper have been done on the supercomputing system
in the Supercomputing Center of Wuhan University
The Project is also funded by China Postdoctoral Science Foundation 2020M682013. F.G.X. is supported by the National SKA Program of China (No. 2020SKA0110102), the NSFC grants 12192223, 12373017, and 12192220. 
Andrzej A. Zdziarski acknowledges support from the Polish National Science Center under the grant 2023/48/Q/ST9/00138.

%=================================================================
%appendix
\appendix
\section{The details of the MCMC approach}
\label{appendix}
The disk's response to a changing accretion rate in its outer region has been investigated under different initial and boundary conditions using the propagating fluctuation model \citep{Lynden-Bell1974MNRAS.168..603L, Pringle1991MNRAS.248..754P, Zdziarski2009MNRAS.399.1633Z,Tanaka2011MNRAS.410.1007T,King1998MNRAS.293L..42K,Lipunova2015ApJ...804...87L}. In our framework, we adopt a semi-infinite disk configuration extending radially to infinity, with a localized fluctuation emerging at a finite radius $R_0$. 

\begin{figure*}
    \centering
    \includegraphics[width=0.8\linewidth]{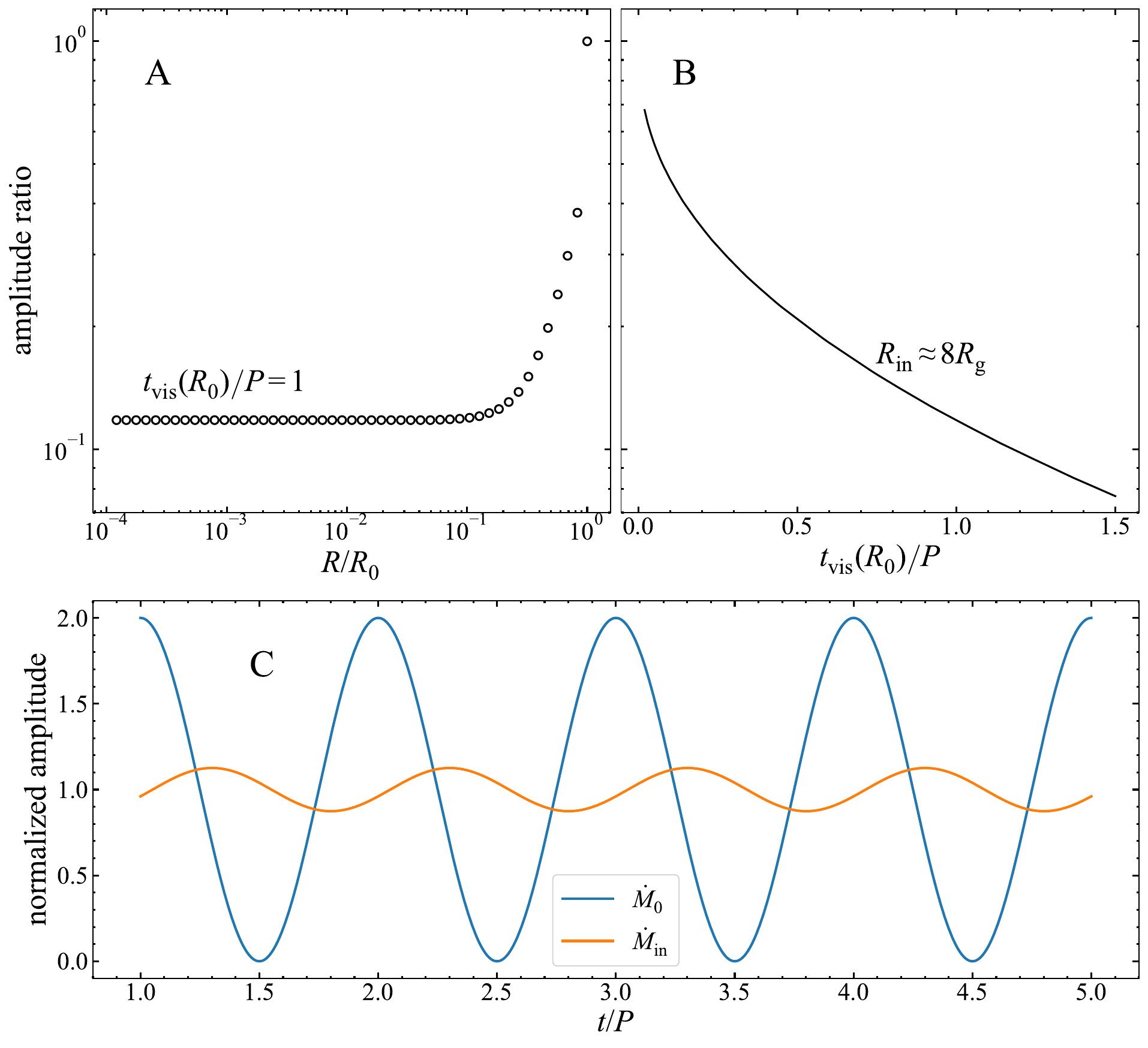}
    \caption{A: The amplitude ratio of the response at different radius to the outer fluctuation for the case of $t_{\rm vis}(R_0)/P=1$; B: The predicted amplitude ratio of the response at $R_{\rm in}\approx 8R_{\rm g}$ to the outer fluctuation for different period. The amplitude of the outer fluctuation source is fixed; C: The temporal evolution of the fluctuation at $R_0$ (blue curve) and $R_{\rm in}$ (orange curve). Note we adopt $n=3/5$ here as an example.}
    \label{fig:greenfun}
\end{figure*}

The resulting accretion rate at each radius in the framework of an infinite disk can be analytically determined using Green's functions if the outer fluctuation $\dot{M}_{0}$ at a given radius $R_0$ is provided \citep{Lynden-Bell1974MNRAS.168..603L,Lyubarskii1997MNRAS.292..679L,Kotov2001MNRAS.327..799K,Zdziarski2009MNRAS.399.1633Z,Mushtukov2018MNRAS.474.2259M}. This can be expressed as
\begin{equation}
    \dot{M}(R,\tau) = \int_0^\infty \dot{M}_0 (R_0,\tau-\tau_1)G_{\dot{M}}(R,\tau_1)\, d\tau_1,
\end{equation}
where $\tau=t/t_{\rm vis}(R_0)$ and $t_{\rm vis}(R_0)$ is the viscous timescale at $R_0$. $G_{\dot{M}}(R,\tau)$ is the Green's function at each radius, which is \citep{Lyubarskii1997MNRAS.292..679L,Zdziarski2009MNRAS.399.1633Z}
\begin{equation}
\begin{aligned}
    G_{\dot{M}}&(R,\tau) =  \frac{\dot{M}_0 2 \mu^2 \xi^{(1/\mu - 1)/2}}{\tau^2} \exp\left[ -\frac{2\mu^2 (\xi^{1/\mu} + 1)}{\tau} \right] \\
    &\times \left\{ {\rm I}_{\mu-1} \left[ \frac{4\mu^2 \xi^{1/(2\mu)}}{\tau} \right] - \xi^{1/(2\mu)} {\rm I}_\mu \left[ \frac{4\mu^2 \xi^{1/(2\mu)}}{\tau} \right] \right\}.   
\end{aligned}
\end{equation}
where ${\rm I}_\mu$ is the modified Bessel function of the first kind, $\xi=\left( R/R_0\right)^{1/2}$, and $\mu=\dfrac{1}{4-2n}$. The value of $n$ is tied to the specific disk model \citep{Shakura1973A&A....24..337S,Zdziarski2009MNRAS.399.1633Z}. The Green's function satisfies $\int_0^\infty G_{\dot{M}}(R,\tau)\text{d}\tau=1$ and $G_{\dot{M}}(R,\tau\le0)\equiv0$. Assuming a periodic outer fluctuation, $\dot{M}_0 =1+ A_0\cos(2\pi t/P)$, at $R_0=10^4 R_{\rm g}$, the response amplitude at different radii depends on both the radius and the fluctuation period. As shown in Figure \ref{fig:greenfun}A, for a specific period (assuming $P=t_{\rm vis}(R_0)$), the fluctuation amplitude declines rapidly once the fluctuation leaves the source and then remains roughly constant as it travels further inward. This result is consistent with the previous analysis \citep{Zdziarski2009MNRAS.399.1633Z}. Furthermore, Figure \ref{fig:greenfun}B shows that, for a fluctuation source with fixed amplitude, viscous damping increases as the period decreases. Thus, the amplitude at $R_{\rm in}$ is naturally suppressed without requiring any intrinsic weakness of the outer fluctuation source. In Figure \ref{fig:greenfun}C, we also plot the temporal evolution of the fluctuation at $R_0$ and $R_{\rm in}$, which indicates that the disk at the interior radius indeed fluctuates on the same period but with a much smaller amplitude than at the outer radius $R_0$.

For the inner disk region, the Green's function at inner radius $R_{\rm in}$ can be simplified as:
\begin{equation}
    G_{\dot{M}}(\tau) = \dfrac{(2\mu^2)^2}{\Gamma (\mu)}\tau^{-1-\mu} \exp \Big(-\dfrac{2\mu^2}{\tau}\Big),
\end{equation}
where $\Gamma(\mu)$ is the Euler Gamma function. Then the inner accretion rate $\dot{M}_{\rm in}$ in response to the outer fluctuation $\dot{M}_0$ is given by:
\begin{equation}
    \dot{M}_{\rm in}(\tau) = \int_0^\infty \dot{M}_0 (\tau-\tau_1)G_{\dot{M}}(\tau_1)\, d\tau_1.
    \label{eq:Min}
\end{equation}
Specifically, we model the accretion rate at a given radius $R_0$ as a sum of three harmonically related cosine functions:
\begin{equation}
    \dot{M}_{0} = \langle \dot{M} \rangle \left[ 1+\sum_{i=1}^{3} A_i \cos \left(\frac{2\pi t}{P_i} + \phi_i\right) \right],\, A_i \in [0,1),
    \label{eq:Mdout}
\end{equation}
for $60204\leq t\leq 60221$. Here, $\langle \dot{M} \rangle$, $A_i$, $P_i$, and $\phi_i$ denote the average accretion rate, fluctuation amplitude, period, and phase, respectively. We assume $P=P_1 = 2P_2 = 4P_3$ to account for harmonic modulation. Further harmonic components are not included, as the purpose of the modeling here is to demonstrate whether the overall temporal morphology and characteristic timescales of the variability are consistent with those expected from propagating accretion-disk fluctuations. For the phases preceding MJD 60204 and following MJD 60221, $\dot{M}_0$ is set to the average rate between MJD 60197 and MJD 60204, and between MJD 60222 and MJD 60223, respectively. This assumption introduces a negligible impact on the results, given the low integrated contributions in these intervals. Using Eq. \ref{eq:Min}, we compute $\dot{M}_{\rm in}$ based on the expression for $\dot{M}_0$ in Eq. \ref{eq:Mdout}.
For simplicity, we assume $L_{\rm disk} = \eta \dot{M_{\rm in}}c^2$, where $\eta$ is assumed to be a constant, given the quasi-stable inner disk radius during the flares. This enables a direct fit of the variation of the disk emission.
More specifically, the variable disk emission is fitted by
\begin{equation}
    L = \int_0^\infty L_{0} (\tau-\tau_1)G_{\dot{M}}(\tau_1)\, d\tau_1,
    \label{eq:damping1}
\end{equation}
\begin{equation}
    L_0 = \langle L \rangle \left[ 1+\sum_{i=1}^{3} A_i \cos \left(\frac{2\pi i }{P}t+\phi_i \right)\right],\, A_i \in [0,1),
    \label{eq:damping2}
\end{equation}
\begin{equation}
    G_{\dot{M}}(\tau) = \dfrac{(2\mu^2)^2}{\Gamma (\mu)}\tau^{-1-\mu} \exp \Big(-\dfrac{2\mu^2}{\tau}\Big),
    \label{eq:damping3}
\end{equation}
where 
\[L_0 = \eta \dot{M}_0 c^2,~ \tau \equiv \frac{t}{t_{\rm vis}(R_0)}\]
\[~P=P_1=2P_2=4P_3, ~\mu=\dfrac{1}{4-2n}.\] 
Here, $L_0$ represents the power of the fluctuation at $R_0$. We adopt $n=3/5$ in the fitting, representing a thin disk supported by gas pressure and dominated by electron scattering \citep{Shakura1973A&A....24..337S,Zdziarski2009MNRAS.399.1633Z}. 

The underlying processes do not operate independently, and we would ideally model the entire light curve using a numerical function. However, in practice, we find this challenging because there are no clear observational clues regarding the phases of the three cosine functions or their temporal evolution. The significant alterations in the modulation of the light curves can be made by even marginal changes in the phase of these functions. Consequently, we are left with the need to segment the light curve and fit each segment with the same functions. We divide the disk flare between MJD 60203 and MJD 60222 into four epochs: MJD 60203-60210, MJD 60210-60215, MJD 60215-60219, and MJD 60219-60222, referred to as $f_2$ to $f_5$. For the Compton flare before MJD 60203, it exhibits different features in the photon index $\Gamma$, $c_f$-$\Gamma$ relation (see Section \ref{subsec:gamma-flux}) and timing properties \citep{Xu2025ApJ...993...40X} compared to the observed in $f_2$-$f_5$, and we refer to this phase as $f_1$.

\begin{figure}
    \centering
    \includegraphics[width=1\linewidth]{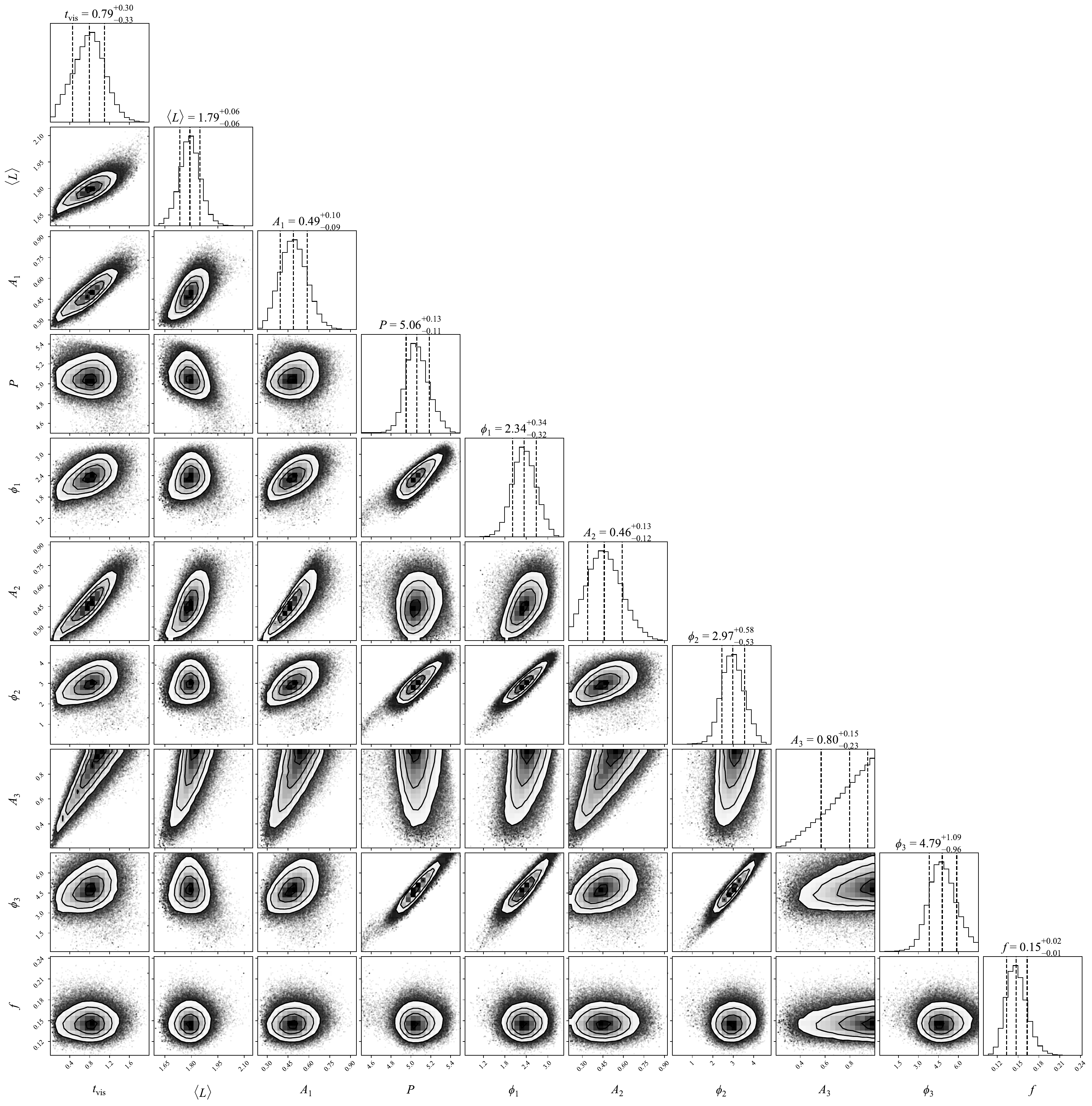}
    \caption{An example corner plot with one and two-dimensional projections of the posterior probability distributions of all the parameters in $f_3$. The second panel from the bottom of each column shows the contours in the two-dimensional projections for every pair of parameters, corresponding to 1-, 2-, and 3-$\sigma$ confidence intervals. The top panel of each column shows the one-dimensional projection of the corresponding parameter. The values above each one-dimensional projection indicate the median value of each parameter, as well as upper and lower limits of 1-$\sigma$ level. The vertical lines in the one-dimensional projections represent the lower, median, and upper values.}
    \label{fig:df1_corner}
\end{figure}

We apply Equations \ref{eq:damping1} to \ref{eq:damping3} to model the disk variability using an MCMC sampler implemented in the {\tt emcee} package in {\tt Python} \citep{Foreman_Mackey_2013}. The maximum likelihood function is defined as \citep{Foreman_Mackey_2013}:
\begin{equation}
\begin{aligned}
    \ln p(y \mid x, \sigma, \theta, f) = -\frac{1}{2} \sum_n \left[ \frac{(y_n - M(x_n, \theta))^2}{s_n^2} + \ln(2\pi s_n^2) \right],
    \label{eq:placeholder_label}
\end{aligned}
\end{equation}
\begin{equation}
    s_n^2=\sigma_n^2+f^2M(x_n,\theta)^2,
\end{equation}
where $M(x_n,\theta)$ is the model prediction, $\sigma_n$ is the data uncertainty. The additional parameter $f$ accounts for a potential uncertainty into the modeling, e.g., the deviation between the cosine approximation and the the true behavior of the fluctuation source. Then the free parameters include $t_{\rm vis}$, $\langle L \rangle$, $A_1$, $P$, $\phi_1$, $A_2$, $\phi_2$, $A_3$, $\phi_3$ and $f$. All the parameters are assigned uniform priors. For the data in $f_5$, we fix $\phi_1=\phi_2=\phi_3$ to reduce the degree due to the limited data points. We initialized an ensemble of 100 walkers randomly distributed around the input values. The sampler was advanced for $N=30000$ steps, with the first 3000 steps discarded as burn-in to ensure chains reach stationarity. Convergence was verified by 
%Gelman-Rubin statistic $\vert \hat{R} - 1 \vert < 0.01$ \citep{Gelman1992StaSc...7..457G} or the 
$N / \tau_{\rm IAT} > 50$ \citep{Goodman2010CAMCS...5...65G}, where $\tau_{\rm IAT}$ is the integrated autocorrelation time (IAT). To mitigate autocorrelation effects, we apply a thinning factor of 20 and flatten the chains to obtain a flat list of samples. This yielded 135,000 independent posterior samples. We then analyze the probability distributions derived from the MCMC chains using {\tt corner} package \citep{Foreman_Mackey_2013}, to estimate the median values and uncertainties quoted at the 1-$\sigma$ level (68\% confidence interval) for each parameter. The example corner plot is presented in Figure \ref{fig:df1_corner}. 

However, we acknowledge that defining the boundaries between epochs involves a degree of subjectivity. To assess the impact of this choice, we conducted the tests by systematically shifting all division boundaries forward and backward 1 day and repeating fit to the disk light curve within the propagating–fluctuation model. In all cases, the model reproduced the overall profile and key features of the observed disk variability, thereby demonstrating that our main physical conclusions regarding fluctuation propagation remain valid.

\clearpage

\bibliography{sample631}{}
\bibliographystyle{aasjournal}

\end{document}